\documentclass[%
preprint,
 amsmath,amssymb,
 aps,12pt
]{revtex4-1}

\usepackage{graphicx}
\usepackage{subfigure}
\usepackage{float}
\usepackage{dcolumn}
\usepackage{bm}
\usepackage{mathrsfs}
\usepackage{mathtools}
\usepackage{lineno}
\usepackage{color,xcolor}
\usepackage{amsmath,amsfonts,amsbsy,amssymb,array,graphicx,psfrag}

\graphicspath{{figures/}}

\newcommand{\EQ}{\begin{equation}}
\newcommand{\EN}{\end{equation}}
\newcommand{\SEQ}{\begin{subequations}}
\newcommand{\SEN}{\end{subequations}}
\newcommand{\EQA}{\begin{eqnarray}}
\newcommand{\ENA}{\end{eqnarray}}
\newcommand{\CS}{\begin{dcases}}
\newcommand{\CN}{\end{dcases}}
\newcommand{\AR}{\begin{array}}
\newcommand{\AN}{\end{array}}
\newcommand\bomg{\boldsymbol{\omega}}
\newcommand{\bs}{\boldsymbol}

\newcommand{\mc}{\mathcal}
\newcommand{\<}{\langle}
\renewcommand{\>}{\rangle}

\newcommand{\rd}{\textrm{d}}
\newcommand{\re}{\textrm{e}}

\newcommand{\lrn}[1]{\left(#1\right)}
\newcommand{\lrl}[1]{\left[#1\right]}

\newcommand{\lra}[1]{\left|#1\right|}

\newcommand{\D}[2]{\frac{\partial #1}{\partial #2}}

\newcommand{\M}[2]{\frac{\textrm{D} #1}{\textrm{D} #2}}

\begin{document}


\title{Effects of the Mach number on the evolution of vortex-surface fields in compressible Taylor--Green flows}

\author{Naifu Peng$^{1,2}$, and Yue Yang$^{1,2,3}$}
\email{yyg@pku.edu.cn}
\affiliation{\small$^{1}$State Key Laboratory for Turbulence and Complex Systems, College of Engineering, Peking University, Beijing 100871, China\\}
\affiliation{\small$^{2}$Center for Applied Physics and Technology, Peking University, Beijing 100871, China\\}
\affiliation{\small$^{3}$Beijing Innovation Center for Engineering Science and Advanced Technology, Peking University, Beijing 100871, China\\}
\date{\today}

\begin{abstract}
  We investigate the evolution of vortex-surface fields (VSFs) in compressible Taylor--Green flows at Mach numbers ($Ma$) ranging from 0.5 to 2.0 using direct numerical simulation.
  The formulation of VSFs in incompressible flows is extended to compressible flows, and a mass-based renormalization of VSFs is used to facilitate characterizing the evolution of a particular vortex surface.
  The effects of the Mach number on the VSF evolution are different in three stages. In the early stage, the jumps of the compressive velocity component near shocklets generate sinks to contract surrounding vortex surfaces, which shrink vortex volume and distort vortex surfaces. The subsequent reconnection of vortex surfaces, quantified by the minimal distance between approaching vortex surfaces and the exchange of vorticity fluxes, occurs earlier and has a higher reconnection degree for larger $Ma$ owing to the dilatational dissipation and shocklet-induced reconnection of vortex lines. In the late stage, the positive dissipation rate and negative pressure work accelerate the loss of kinetic energy and suppress vortex twisting with increasing $Ma$.

  \begin{description}
  \item[Keywords]
  Compressible Taylor-Green flow, shocklet-vortex interaction, vortex reconnection
  \end{description}
\end{abstract}

\maketitle


\section{Introduction}
\label{sec:introduction}

The effects of the Mach number ($Ma$) or compressibility on vortex dynamics are important for the investigation of compressible flows \cite{Smith1986}, such as compressible hydrodynamic instability \cite{Brouillette2002} and turbulence \cite{Samtaney2001,Wang2012}.
In general, the vortex dynamics in compressible flows, in which the velocity field and thermodynamic properties are coupled \cite{Shapiro1953}, is more complex than that in incompressible flows owing to the strong nonlinear interactions of vortices and compressing processes such as shock waves \cite{Wu2015}.

In early studies, the compressibility effects on the geometry of simple vortical structures were mainly investigated using experiments and theoretical analysis \cite{Mack1960,Brown1965,Colonius1991}.
In recent decades, the Mach-number effects on the dynamics of complex vortical structures were studied facilitated by the development of numerical simulation.
Moore and Pullin \cite{Moore1987} presented that the compressibility effects reduce the speed of propagation of a vortex pair and distort vortex boundaries using the hodograph-plane numerical method. This method was then applied to the study of vortex arrays in a shock-free shear layer, and showed that the vortices shrink in size and become closer with increasing $Ma$ \cite{Ardalan1995}. Sandham \cite{Sandham1994} demonstrated that the compressibility effects delay the vortex pairing process and elongate vortex trajectories in a mixing layer using direct numerical simulation (DNS).

In particular, some important processes of compressible vortex dynamics, such as the shocklet-vortex interaction and the compressibility effects on vortex reconnection, can be investigated in detail using DNS. Virk \emph{et al.} \cite{Virk1995} studied compressible vortex reconnection of two antiparallel vortex tubes, and found that the shocklets induce the earlier occurrence of the vortex reconnection in larger $Ma$ flows, and reduce the reconnection degree and curvature of isosurfaces of the vorticity magnitude.
Hickey \emph{et al.} \cite{Hickey2016a} found that the compressibility effects inhibit the streamwise communication and increase the cross-wake communication of neighboring vortices in transitional high-speed
planar wakes. In compressible turbulence, a series of studies of Lee \emph{et al.} \cite{Lee1991,Lee1993,Lee1997} showed that shocklets change the vorticity and dilatation distribution in high-$Ma$ turbulent flows, which leads to strong shocklet-vortex interactions.

However, most of the existing studies identify the `vortex cores' using the Eulerian vortex-identification criteria such as the vorticity magnitude and the others based on the local velocity gradient tensor. These methods can identify nonunique vortical structures owing to the subjective selection of the isocontour level, and the structures identified at different times do not have strong temporal coherence in principle.

To tackle the long-standing issue of characterizing the evolution of vortical structures, Yang and Pullin \cite{Yang2010,Yang2011} developed the vortex-surface field (VSF), whose isosurfaces are vortex surfaces consisting of vortex lines. This method is rooted in the Helmholtz vorticity theorem, but it can describe the Lagrangian-like evolution of vortex surfaces in viscous flows with numerical regularization.
The VSF method has been applied to incompressible viscous highly-symmetric flows \cite{Yang2011}, and transitional channel flow \cite{Zhao2016b,Xiong2017} and boundary layer \cite{Yang2016}. Numerical results on the evolution of VSFs clarify the continuous vortex dynamics in these transitional flows including vortex reconnection, rolling-up of vortex tubes, vorticity intensification, and vortex stretching and twisting from a Lagrangian perspective in terms of topology and geometry of vortex surfaces.

In order to extend the capability of the VSF, the compressible Taylor--Green (TG) flow is a good candidate for the first study of the evolution of VSFs in compressible flows, because the TG flow has the symmetries facilitating to construct exact initial VSFs, and it is a transitional flow with the major vortex dynamics such as vortex stretching, reconnection, rolling-up, and twisting. The incompressible TG flow was proposed and analyzed by Taylor and Green \cite{Taylor1937}, and investigated in detail by Brachet \emph{et al.} \cite{Brachet1983,Brachet1984} using DNS. Shu \emph{et al.} \cite{Shu2002} added the weak compressibility to the TG flow to verify the convergence of numerical schemes. Then the compressible TG flow has become a popular benchmark case for comparing accuracy and efficiency of numerical schemes \cite{Drikakis2007,Chapelier2012,Debonis2013,Shirokov2014,Bo2017} in the last decade. On the other hand, the Mach numbers are relatively low as the order of $10^{-1}$ in the existing studies, so the vortex dynamics in the weakly compressible TG flows is not distinguished from that in incompressible TG flows.

In the present study, we apply the VSF to investigate the compressibility effects on the vortex dynamics in compressible TG flows at a range of Mach numbers from 0.5 to 2.0.
First, we extend the formulation of the VSF method in incompressible flows to compressible flows.
Then we address three specific issues in compressible TG flows:
(1) how to elucidate the interaction mechanism of the generation of shocklets and the deformation of vortex surfaces;
(2) how the compressibility affect the process of vortex reconnection;
(3) how to explain the difference of the geometries of vortex tubes in low- and high-$Ma$ flows in the late, nearly turbulent stage.

The outline of this paper is as follows. In Sec.~\ref{sec:VSF}, we describe the VSF method in compressible flows. In Sec.~\ref{sec:num}, we present the numerical method for the DNS of compressible TG flows and validate the effectiveness of the two-time method for solving the VSF equations. In Sec.~\ref{sec:results}, we quantitatively investigate the effects of the Mach number on the evolution of VSFs in different stages. Some conclusions are drawn in Sec.~\ref{sec:conclusion}.

\section{VSF method in compressible flows}
\label{sec:VSF}

\subsection{Governing equations for compressible flows}
\label{sec:VSF_NS}

The compressible flow of ideal gas without body force is governed by the three-dimensional compressible Navier--Stokes (N--S) equations
\EQ
\label{eq:ns}
\CS
\D{\bs{U}}{t}+\D{\bs{F}_j}{x_j}-\D{\bs{V}_j}{x_j}=0,\\
p=\rho RT,
\CN
\EN
where the vector-valued quantities are
\EQ
\bs{U}=
\lrn{\AR{c}
\rho\\
\rho u_1\\
\rho u_2\\
\rho u_3\\
E
\AN},~~
\bs{F}_j=
\lrn{\AR{c}
\rho u_j\\
\rho u_1u_j+p\delta_{1j}\\
\rho u_2u_j+p\delta_{2j}\\
\rho u_3u_j+p\delta_{3j}\\
(E+p)u_j
\AN},~~
\bs{V}_j=
\lrn{\AR{c}
0\\
\sigma_{1j}\\
\sigma_{2j}\\
\sigma_{3j}\\
\sigma_{jk}u_k+\kappa\D{T}{x_j}
\AN}.
\EN
Here, the subscript $j = 1,2,3$ denotes indices in three-dimensional Cartesian coordinates, i.e., $x_1$, $x_2$, and $x_3$ are equivalent to $x$, $y$, and $z$, respectively, and $u_1$, $u_2$, and $u_3$ are equivalent to $u$, $v$, and $w$, respectively; $\delta_{ij}$ is the Kronecker delta function, and the Einstein summation convention is applied; $\rho$ is the density, $p$ is the static pressure, $\kappa$ is the thermal conductivity, $T$ is the temperature, and $R$ is the universal gas constant. Under the perfect gas assumption, the total energy is
\EQ
E=\frac{p}{\gamma-1}+\frac{1}{2}\rho u_j u_j,
\EN
where $\gamma\equiv C_p/C_v$ is the adiabatic exponent of gas as the ratio of the heat capacity $C_p$ at constant pressure to the heat capacity $C_v$ at constant volume. For compressible Newtonian flows, the viscous stress tensor is
\EQ
\sigma_{ij}=\mu\lrn{\D{u_i}{x_j}+\D{u_j}{x_i}}-\frac{2}{3}\mu\theta\delta_{ij},
\EN
where $\theta\equiv\nabla\cdot\bm{u}$ is the dilatation, and the dynamic viscosity $\mu$ is assumed to obey Sutherland's law \cite{Sutherland1893}
\EQ
\mu=\frac{1.4042T^{1.5}}{T+0.40417}\mu_\infty
\EN
with $\mu_\infty=1.716\times 10^{-5}$kg/(m$\cdot$s).

The evolution equation of the vorticity $\bomg\equiv\nabla\times\bs{u}$ in compressible flows \cite{Wu2015} is
\EQ
\label{eq:omg}
\M{\bm{\omega}}{t}=\left(\bm{\omega}\cdot\nabla\right)\bm{u}-\theta\bm{\omega}+\frac{1}{\rho^2}\lrn{\nabla\rho\times\nabla p}+\nu\nabla^2\bm{\omega},
\EN
where the material derivative is defined as
\EQ
\M{}{t}\equiv\D{}{t}+\bm{u}\cdot\nabla
\EN
and $\nu\equiv\mu/\rho$ is the kinematic viscosity. The dilatational term $\theta\bm{\omega}$ and baroclinic term $(\nabla\rho\times\nabla p)/\rho^2$ in Eq.~\eqref{eq:omg} can create vorticity in compressible flows, while they are vanishing in incompressible flows.

\subsection{Evolution equations of the VSF}
\label{sec:VSF_evolution}

The VSF $\phi_v$ is defined to satisfy the constraint $\bomg\cdot\nabla\phi_v=0$, so that every isosurface of $\phi_v$ is a vortex surface consisting of vortex lines.
The evolution equations of the VSF have been derived in incompressible flows \cite{Yang2010}, and we extend the derivation to compressible flows.
Based on Eq.~\eqref{eq:omg} and the constraint equation for the VSF~\cite{Yang2010}
\EQ
\M{}{t}(\bomg\cdot\nabla\phi_v)=0,
\EN
we obtain the evolution equations of the VSF for viscous compressible flows as 
\EQ
\label{eq:CVSF}
\CS
\M{\phi_v}{t}=\nu\mc L_\nu+\mc L_c,\\
\lrl{\nu\nabla^2\bm{\omega}+\frac{1}{\rho^2}\lrn{\nabla\rho\times\nabla p}}\cdot\nabla\phi_v+\bm{\omega}\cdot\nabla\lrn{\nu\mc L_\nu+\mc L_c}=0,
\CN
\EN
after some algebra, where $\mc L_\nu(\bm{x},t)$ and $\mc L_c(\bm{x},t)$ are assumed to be regular scalar fields, and they represent the effects of viscosity and baroclinicity on the evolution of VSFs, respectively.

It is noted that Eq.~\eqref{eq:CVSF} can be simplified to the VSF equations in incompressible barotropic flows with vanishing $\nabla\rho\times\nabla p$ and $\mc L_c$ as \cite{Yang2010}
\EQ
\label{eq:IVSF}
\CS
\M{\phi_v}{t}=\nu\mc L_\nu,\\
\nabla^2\bm{\omega}\cdot\nabla\phi_v+\bm{\omega}\cdot\nabla\mc L_\nu=0.
\CN
\EN
Then for inviscid flow with vanishing $\nu$, Eq.~\eqref{eq:IVSF} can be further simplified as
\EQ\label{eq:IIVSF}
\M{\phi_v}{t}=0,
\EN
where $\phi_v$ remains a Lagrangian scalar field in the temporal evolution and Eq.~\eqref{eq:IIVSF} represents the Helmholtz vorticity theorem \cite{Helmholtz1858,Yang2010}.

The baroclinic term $(\nabla\rho\times\nabla p)/\rho^2$  in Eq.~\eqref{eq:CVSF} contains the effect of compressibility on the evolution of VSFs, and the dilatation--vorticity term $\theta\bm{\omega}$ in Eq.~\eqref{eq:omg} has no effect on the VSF evolution. Therefore, if we track the Lagrangian scalar as Eq.~\eqref{eq:IIVSF} in a compressible viscous flow with the initial field as a VSF,
both viscous and baroclinic effects can drive the scalar field to deviate from the desired VSF solution, which corresponds to the violation of the Helmholtz theorem.

Additionally, the existence and uniqueness of the solution of Eq.~\eqref{eq:CVSF} are not satisfied \cite{Huang1997,Yang2010}, so appropriate numerical regularization must be introduced to restore the uniqueness of the VSF solution, which will be discussed in Sec.~\ref{sec:num_impl}.

\section{Numerical methods}
\label{sec:num}

\subsection{DNS of compressible TG flows}
\label{sec:num_DNS}

In the numerical implementation, Eq.~\eqref{eq:ns} is non-dimensionalized by the characteristic length $L$, velocity $U$, time $L/U$, density $\rho_\infty$, temperature $T_\infty$, pressure $\rho_\infty RT_\infty$, kinetic energy $\rho_\infty U^2$, sound speed $c_\infty=\sqrt{\gamma RT_\infty}$, dynamic viscosity $\mu_\infty$, and heat conductivity coefficient $\kappa_\infty$. The dimensionless variables with the superscript asterisk are introduced as
\EQ
\label{eq:nond}
\begin{split}
&\bm{x}=\bm{x}^*L,~~\bm{u}=\bm{u}^*U,~~t=t^*\frac{L}{U},~~\rho=\rho^*\rho_\infty,~~T=T^*T_\infty,\\
&p=p^*\rho_\infty U^2,~~E=E^*\rho_\infty U^2,~~c=c^*c_\infty,~~\mu=\mu^*\mu_\infty,~~\kappa=\kappa^*\kappa_\infty.
\end{split}
\EN
Substituting Eq.~\eqref{eq:nond} into Eq.~\eqref{eq:ns} and omitting all the asterisks for clarity yield the dimensionless form of compressible N--S equations \cite{Samtaney2001}
\EQ
\label{eq:nsd}
\CS
\D{\rho}{t}+\D{\lrn{\rho u_j}}{x_j}=0,\\
\D{\lrn{\rho u_i}}{t}+\D{}{x_j}\lrn{\rho u_iu_j+\frac{p}{\gamma Ma^2}\delta_{ij}}=\frac{1}{Re}\D{\sigma_{ij}}{x_j},\\
\D{E}{t}+\D{}{x_j}\lrl{\lrn{E+\frac{p}{\gamma Ma^2}}u_j}=\frac{1}{\alpha}\D{}{x_j}\lrn{\kappa\D{T}{x_j}}+\frac{1}{Re}\D{\lrn{\sigma_{ij}u_i}}{x_j},\\
p=\rho T,
\CN
\EN
with $\alpha=PrRe(\gamma-1)Ma^2$ and dimensionless viscous stress tensor
\EQ
\sigma_{ij}=\mu\lrn{\D{u_i}{x_j}+\D{u_j}{x_i}}-\frac{2}{3}\mu\theta\delta_{ij}
\EN
and total energy
\EQ
\label{eq:e}
E=\frac{p}{(\gamma-1)\gamma Ma^2}+\frac{1}{2}\rho u_j u_j.
\EN
We remark that all the quantities are in the dimensionless form in following expressions.
Thus Eq.~\eqref{eq:nsd} is governed by four dimensionless parameters, including the Reynolds number $Re\equiv\rho_\infty UL/\mu_\infty$, Mach number $Ma\equiv U/c_\infty$, Prandtl number $Pr\equiv\mu_\infty C_p/\kappa_\infty$, and adiabatic exponent of gas $\gamma$. We set $Pr=0.7$ and $\gamma=1.4$, and the system is then only affected by $Re$ and $Ma$.

The TG flow is a simple transitional flow in a three-dimensional periodic box \cite{Taylor1937}. The periodic box lies in $0\le x,y,z\le 2\pi$ and the initial solenoidal velocity field \cite{Brachet1983} is
\EQ
\label{eq:TG2}
\CS
u_0=\sin x\cos y\cos z,\\
v_0=-\cos x\sin y\cos z,\\
w_0=0,
\CN
\EN
where the subscript `0' denotes the quantity at the initial time $t=0$. 
The initial pressure
\EQ
p_0=1+\frac{1}{16}\left(\cos 2x+\cos 2y\right)\left(\cos 2z+2\right)
\EN
is calculated from the pressure Poisson equation with Eq.~\eqref{eq:TG2}. The initial temperature is set as $T_0=1$, and the initial density is $\rho_0=p_0/T_0$.

In the DNS of compressible TG flows by solving Eq.~\eqref{eq:nsd}, the convective terms are discretized by the hybrid compact eighth-order finite difference and seventh-order weighted essentially non-oscillatory (FD--WENO) scheme with hyperviscosity, the diffusion terms are discretized by the compact eighth-order finite difference scheme \cite{Wang2010}, and the time integration is advanced by the third-order total-variation-diminishing (TVD) Runge--Kutta scheme \cite{Gottlieb1998}. These numerical schemes have been validated to be stable and reasonably accurate for the DNS of compressible isotropic turbulence at high $Ma$ \cite{Wang2010}.

In the present study, the numerical cases of compressible TG flows are chosen at a moderate Reynolds number $Re=400$, which is the same as that in the former study of VSFs in incompressible TG flows \cite{Yang2011}, and at four Mach numbers $Ma=0.5$, 1.0, 1.5, and 2.0 varying from weak to strong compressibility. The DNS of Eq.~\eqref{eq:nsd} in the periodic box is on uniform grids $N^3$, and in a time period from $t=0$ to $t_{\max}=10$ with the time step $\Delta t = 5\times 10^{-4}$ satisfying the Courant--Friedrichs--Lewy (CFL) condition.

\begin{figure}
\begin{center}
\includegraphics[width=6.2in]{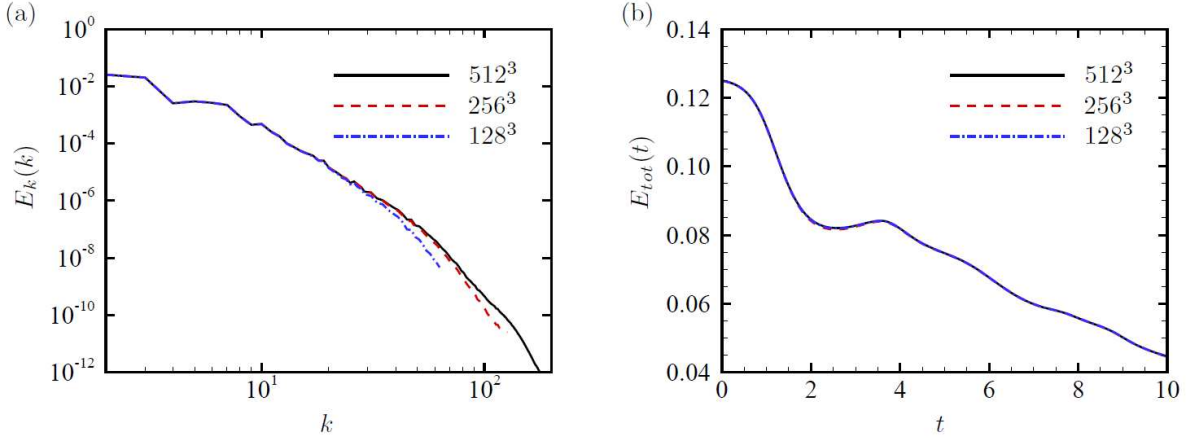}
\caption{Statistics of the kinetic energy for $Ma=2.0$ with three gird resolutions. (a) Kinetic energy spectra at $t=7$; (b) The temporal evolution of the total kinetic energy.}\label{fig:convergence}
\end{center}
\end{figure}

In order to check the grid convergence, we set the Mach number at the highest value $Ma=2.0$ and test the DNS on three different numbers of grids $N^3 = 128^3$, $256^3$, and $512^3$.
Flow statistics of the three runs are compiled in Table~\ref{tab:grid_test}, where the viscous dissipation $\varepsilon$ is defined as
\EQ
\label{eq:epsilon}
\varepsilon\equiv\frac{1}{Re}\<\sigma_{ij}S_{ij}\>=\frac{1}{Re}\<\mu|\bomg|^2\>+\frac{4}{3Re}\<\mu\theta^2\>.
\EN
For an arbitrary quantity $f$, $\<f\>$ denotes the volume averaging of $f$, $f'$ denotes the root-mean-square (rms) of $f$ in the volume, and $\overline{f}$ denotes the time averaged value from $t=0$ to $t=t_{\max}$.
Furthermore, the kinetic energy spectra $E_k(k)$ at $t=7$, where $k$ is the wavenumber, and the temporal evolution of kinetic energy $E_{tot}(t)$ with the three grid resolutions are plotted in Fig.~\ref{fig:convergence}.
We find that all the statistics in Table~\ref{tab:grid_test} and Fig.~\ref{fig:convergence} are converged at the grids $N^3=512^3$ with increasing the grid resolution.

\begin{table}
\caption{\label{tab:grid_test}Statistics in the assessment of grid convergence at $Ma=2.0$.}
\begin{ruledtabular}
\begin{tabular}{ccccc}
Grids & $\overline{u'}$ & $\overline{\omega'}$ & $\overline{\varepsilon}$ & $\overline{\rho'}$\\
\colrule
$128^3$    & 0.229           & 1.132                & 0.0054                   & 0.274\\
$256^3$    & 0.229           & 1.134                & 0.0060                   & 0.275\\
$512^3$    & 0.229           & 1.134                & 0.0063                   & 0.275\\
\end{tabular}
\end{ruledtabular}
\end{table}

\subsection{Two-time method for calculating VSFs}
\label{sec:num_impl}

Owing to the non-existence and non-uniqueness of the VSF solution of Eq.~\eqref{eq:CVSF} \cite{Huang1997,Yang2010}, the two-time method has been developed to solve Eq.~\eqref{eq:CVSF} with numerical regularization, which is described in Ref.~\cite{Yang2011} in detail.
In this method, each time step is divided into prediction and correction steps.
In the prediction step, the temporary VSF solution $\phi_v^*(\bs{x},t)$ that is evolved as a Lagrangian field is calculated in physical time $t$ as
\EQ
\label{eq:phy}
\D{\phi_v^*(\bs{x},t)}{t}+\bs{u}(\bs{x},t)\cdot\nabla\phi_v^*(\bs{x},t)=0,~~t\ge 0,
\EN
where $\bs u(\bs{x},t)$ is the Eulerian velocity obtained from DNS, and the temporary $\phi_v^*(\bs{x},t)$ can slightly deviate from an accurate VSF at each physical time step owing to the breakdown of the Helmholtz vorticity theorem in compressible viscous flows.

Then in the correction step, at a fixed physical time $t$, the temporary VSF $\phi_v^*(\bs{x},t)$ is projected on the desired accurate VSF solution $\phi_v(\bs{x},t;\tau)$ and it is transported along the frozen vorticity $\bomg(\bs{x},t)$ in pseudo-time $\tau$ as
\EQ
\label{eq:pse}
\D{\phi_v(\bs{x},t;\tau)}{\tau}+\bomg(\bs{x},t)\cdot\nabla\phi_v(\bs{x},t;\tau)=0,~~0\le\tau\le T_\tau,
\EN
where the initial condition is $\phi_v(\bs{x},t;\tau=0)=\phi_v^*(\bs{x},t)$. Finally, $\phi_v$ is updated by $\phi_v(\bs{x},t;\tau=T_\tau)$ from Eq.~\eqref{eq:pse} for each physical time step, where $T_\tau$ is the maximal value of the pseudo-time when $\phi_v(\bs{x},t;\tau)$ is converged in Eq.~\eqref{eq:pse}.

In the numerical implementation of Eqs.~\eqref{eq:phy} and \eqref{eq:pse}, integrations in $t$ and $\tau$ are marched by the TVD Runge--Kutta scheme \cite{Gottlieb1998}, and the pseudo-time step satisfies the CFL condition based on $\bomg$ \cite{Yang2011}. The convection terms are approximated by the fifth-order WENO scheme \cite{Jiang1996}. The numerical diffusion in the WENO scheme serves as a numerical dissipative regularization for removing small-scale, nearly singular scalar structures in Eqs.~\eqref{eq:phy} and \eqref{eq:pse}. The initial VSF \cite{Yang2011} for TG flows is
\EQ
\label{eq:vsftg}
\phi_{v0}=\left(\cos 2x-\cos 2y\right)\cos z
\EN
that satisfies the VSF constraint $\bomg_0\cdot\nabla\phi_{v0}=0$. Periodic boundary conditions are applied for Eqs.~\eqref{eq:phy} and \eqref{eq:pse}. The grids $512^3$ for VSFs are the same as those of DNS, and TG symmetries are used to significantly reduce the computational cost for VSFs \cite{Yang2010}.

\begin{figure}
\begin{center}
\includegraphics[width=6.2in]{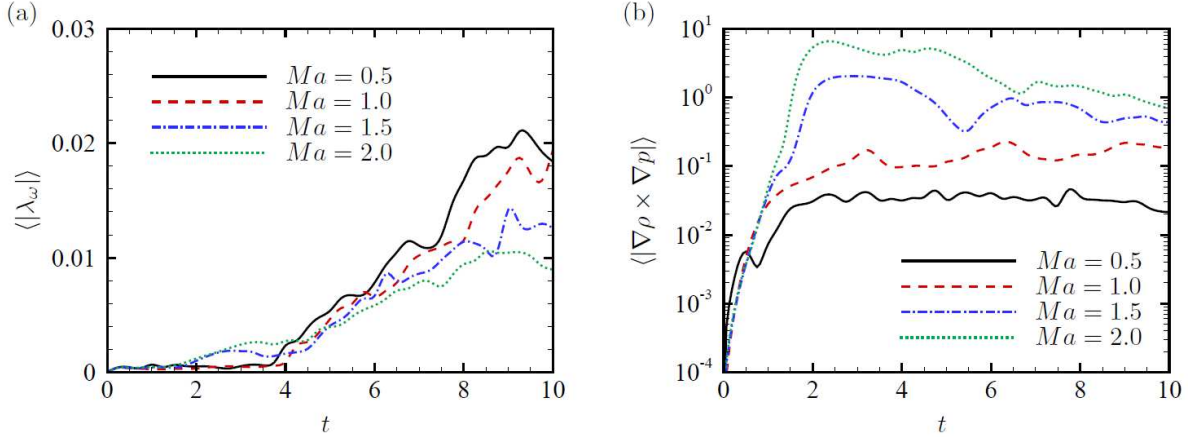}
\caption{The error assessment of the VSF method in compressible TG flows at different $Ma$. (a) Temporal evolution of the volume averaged VSF deviation; (b) Temporal evolution of the volume averaged absolute value of the baroclinic term in the vorticity Eq.~\eqref{eq:omg}.}\label{fig:div}
\end{center}
\end{figure}

In general, the numerical VSF solution $\phi_v$ has a slight deviation from the exact VSF, and the deviation is defined by the cosine of the angle between the vorticity and the gradient of VSF solution as \cite{Yang2010}
\EQ
\lambda_\omega\equiv\frac{\bomg\cdot\nabla\phi_v}{|\bomg||\nabla\phi_v|}.
\EN
In different incompressible transitional flows \cite{Yang2011,Zhao2016b}, it has been shown that 
$\<|\lambda_\omega|\>$
can be controlled below 5\% in the temporal evolution of VSFs from theoretical approximation and numerical experiments.
In the present compressible TG flows, the temporal evolution of $\<\lra{\lambda_{\omega}}\>$ and the volume-averaged baroclinic term that can cause the violation of the Helmholtz theorem in Eq.~\eqref{eq:CVSF} are plotted in Fig.~\ref{fig:div} with different Mach numbers.
Although the baroclinic effect becomes notable with increasing $Ma$, as the magnitude of $\<|\nabla\rho\times\nabla p|\>$ is $O(1)$ at $Ma=2.0$, $\<\lra{\lambda_{\omega}}\>$ is still less than $3\%$. Hence, the two-time method can effectively reduce the VSF deviation to achieve reasonably accurate VSF solutions in compressible flows at high $Ma$.

In addition, the isosurfaces of $\phi_v$ with the same isocontour level at different times should have strong time coherence to display the temporal evolution of a vortex surface.
However, the boundness of $\phi_v$ in the numerical implementation is not ensured for long times owing to the numerical dissipative regularization in the two-time method. Thus we renormalize the VSF for postprocessing, so that
the mass of the fluid enclosed in each isosurface of the normalized VSF is enforced to be conserved during the temporal evolution.
This renormalization is consistent with the Helmholtz theorem, and can be a good approximation for the flow conditions that are not very far from the requirement of the Helmholtz theorem.

The normalized VSF $\hat{\phi}_v$ is implicitly determined from
\EQ
\hat{M}(\hat{\phi}_v = \varphi,t) = \hat{M}(\phi_{v0} = \varphi,t=0)
\EN
by searching the isocontour values $\hat{\phi}_v=\varphi$ at a given time $t$ and $\phi_{v0}$ at the initial time which correspond to the same fluid mass. Here, the fluid mass within an isosurface of $\phi_v = \varphi$ is calculated as
\EQ
\label{eq:mass}
\hat{M}(\phi_v = \varphi)=\int\rho H_\epsilon(\phi_v = \varphi)\rd V,
\EN
and the discretized Heaviside function \cite{Chang1996} is
\EQ
H_\epsilon(\phi_v = \varphi)=
\CS
0,&\phi_v-\varphi<-\epsilon,\\
\frac{1}{2\epsilon}\lrn{\phi_v+\epsilon}+\frac{1}{2\pi}\sin\frac{\pi\phi_v}{\epsilon}, &\lra{\phi_v-\varphi}\le\epsilon,\\
1,&\phi_v-\varphi>\epsilon,
\CN
\EN
with a smoothing parameter $\epsilon=1\times 10^{-6}$.
In numerical experiments, we find that the isosurface of $\hat{\phi}_v$ with a given isocontour value can reasonably describe the continuous temporal evolution of a particular vortex surface.

\section{Results and discussion}
\label{sec:results}

\begin{figure}
\begin{center}
\includegraphics[width=5.5in]{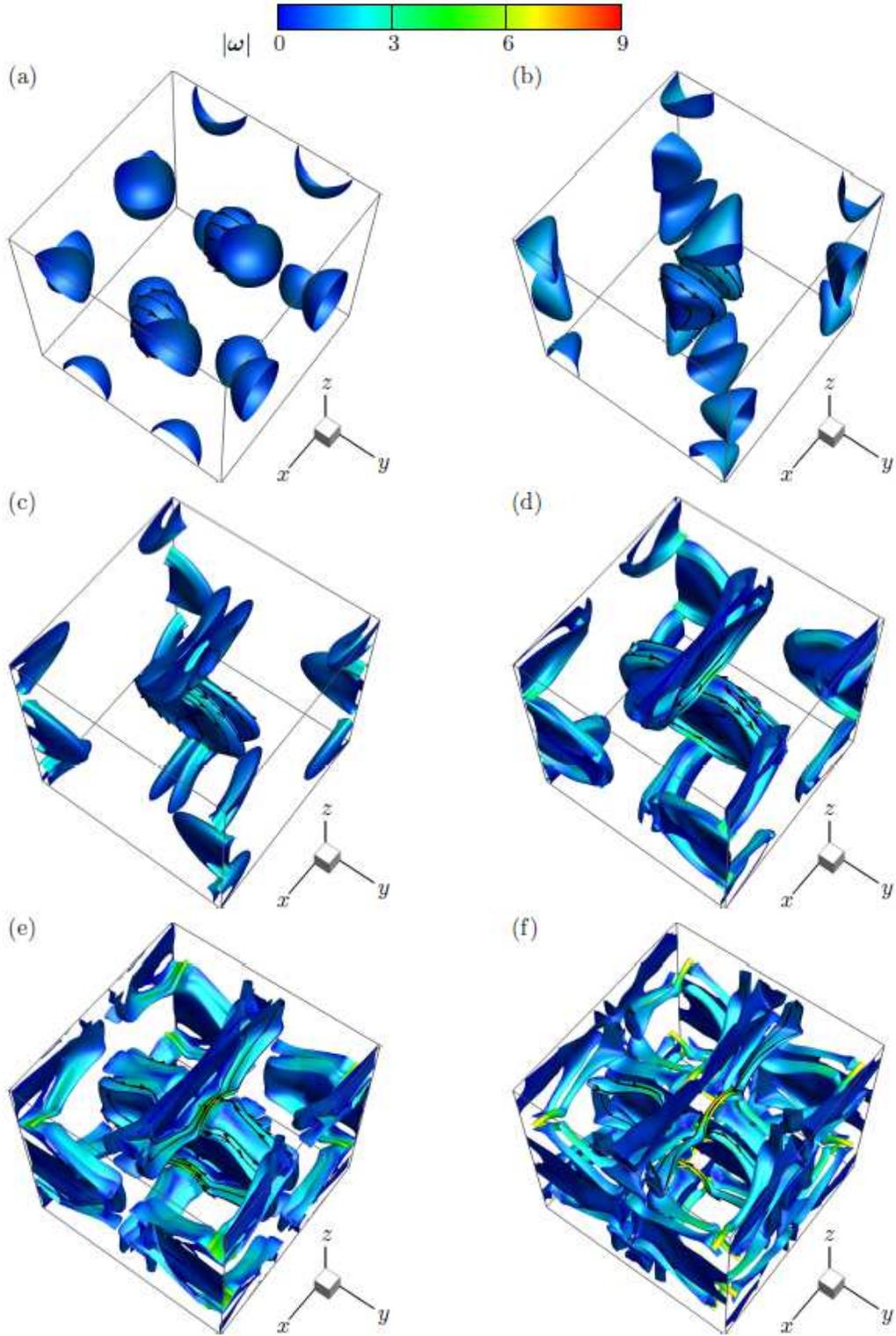}
\caption{Evolution of the VSF isosurface of $\hat{\phi}_v=0.5$ in the TG flow at $Ma=2.0$. Some vortex lines are integrated and plotted on the isosurfaces. The surfaces are color coded by $0\le|\bomg|\le 9$ from blue to red. (a) $t=0$, (b) $t=1.5$, (c) $t=3$, (d) $t=4.5$, (e) $t=6$, (f) $t=7$.}\label{fig:2.0phi}
\end{center}
\end{figure}

The temporal evolution of isosurfaces of the renormalized VSF $\hat{\phi}_v=0.5$ in the TG flow at $Ma=2.0$ is shown in Fig.~\ref{fig:2.0phi}. The extracted vortex surfaces are color coded by $0\le|\bomg|\le 9$ from blue to red. Some vortex lines are integrated from points on the isosurfaces, and we can see that they almost lie on the surfaces, which agrees with the small VSF deviation in the simulation.
From the observation in Fig.~\ref{fig:2.0phi}, we roughly divide the entire evolution of VSFs into three stages based on topological and geometrical features of vortex surfaces: (1) vortex deformation in the early stage in Figs.~\ref{fig:2.0phi} (a) and (b); (2) vortex reconnection during transition in Figs.~\ref{fig:2.0phi} (c) and (d); (3) rolling-up, stretching, and twisting of vortex tubes in the late stage in Figs.~\ref{fig:2.0phi} (e) and (f).

The visualizations of the temporal evolution of VSF isosurfaces at lower $Ma$ and in incompressible TG flows \cite{Yang2011} also show the three stages in the VSF evolution, but they have notable differences in the quantified geometry and topology owing to the Mach number effects on vortex evolution, which will be discussed in detail in this section.

In addition, the evolution of the isosurfaces of $|\bomg|$ in compressible TG flows only have slight geometric differences (not shown) from those in incompressible TG flows \cite{Brachet1983}.
Yang and Pullin (see Figs.~5, 8, and 9 in Ref.~\cite{Yang2011}) demonstrated that the isosurfaces of $|\bomg|$ and other Eulerian vortex-identification criteria are very different from vortex surfaces owing to the significant misalignment of the isosurfaces and nearby vortex lines, and the evolution of the vortical structures identified by Eulerian vortex identification criteria cannot be elucidated by the Helmholtz vorticity theorem.
Therefore, the Eulerian vortical structures identified at different times are hard to reveal the Lagrangian-like, continuous evolution of a particular vortex surface, and characterize vortex reconnection in TG flows.

\subsection{Vortex deformation in the early stage}
\label{sec:results_early}

\begin{figure}
\begin{center}
\includegraphics[width=6.2in]{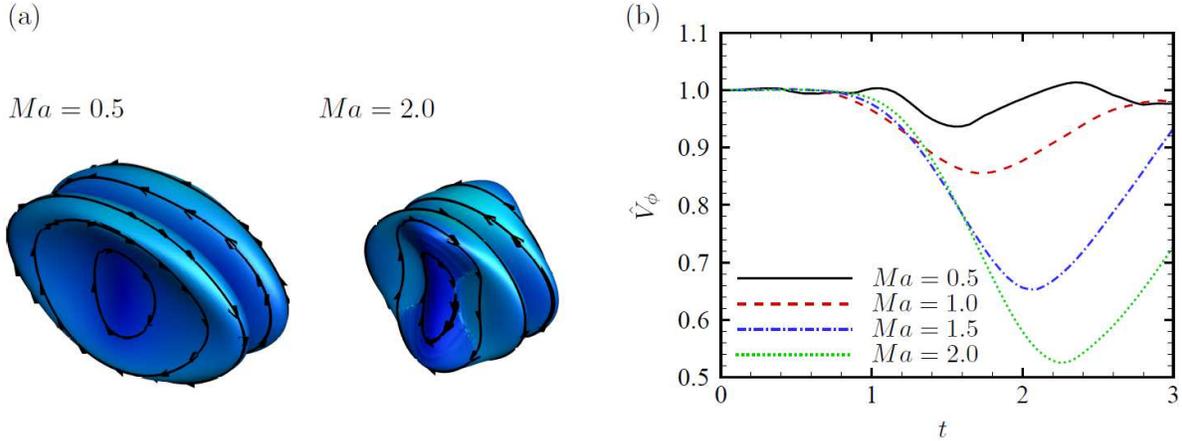}
\caption{Comparison of the volume of vortex surfaces at different $Ma$. (a) Visualization of the isosurfaces of $\hat{\phi}_v=0.5$ at $t=2.2$ for $Ma=0.5$ and 2.0. The surfaces are color coded by $0\le|\bomg|\le 9$ from blue to red; (b) The temporal evolution of the renormalized volume $\hat{V}_\phi$ at different $Ma$.}\label{fig:def}
\end{center}
\end{figure}

In the early stage of the VSF evolution during the time period of $0\le t<t^*$, where $t^*$ is the estimated time when the vortex reconnection occurs, vortex surfaces only have geometrical deformation without topological changes.
Considering the symmetries in TG flows \cite{Brachet1983}, we extract the minimum subdomain within $\pi/2\le x,y,z\le 3\pi/2$ to visualize vortex surfaces for $Ma=0.5$ and 2.0 at $t=2.2$ in Fig.~\ref{fig:def}(a).
We find that the geometry of the vortex surfaces of $Ma=0.5$ is almost identical to that in the incompressible TG flow \cite{Yang2011}, but the vortex surfaces of $Ma=2.0$ appear to be squeezed by the compressibility effect.

In order to quantify the deformation of vortex surfaces, we calculate the volume change of the fluid enclosed in closed vortex surfaces, and plot the temporal evolution of the normalized volume $\hat{V}_\phi\equiv V_\phi(\hat{\phi}_v = \varphi)/V_0(\hat{\phi}_{v0} = \varphi)$ at different Mach numbers in Fig.~\ref{fig:def}(b).
Here, $V_\phi = \int H_\epsilon(\hat{\phi}_v = \varphi)\rd V$ is the volume of the fluid enclosed in the isosuface of $\hat{\phi}_v=\varphi$, and $V_0$ is the initial volume of the fluid enclosed in the isosurface of $\hat{\phi}_{v0}=\varphi$. For a VSF isosurface, $\hat{V}_\phi < 1$ characterizes the volume shrinkage of a vortex surface in compressible flows. We find that $\hat{V}_\phi$ is decreased with increasing $Ma$ for a typical VSF isosurface of $\varphi = 0.5$, and $\hat{V}_\phi$ reaches the minimum at $t=2.2$ for $Ma=2.0$ with strong compressibility.

\begin{figure}
\begin{center}
\includegraphics[width=6.0in]{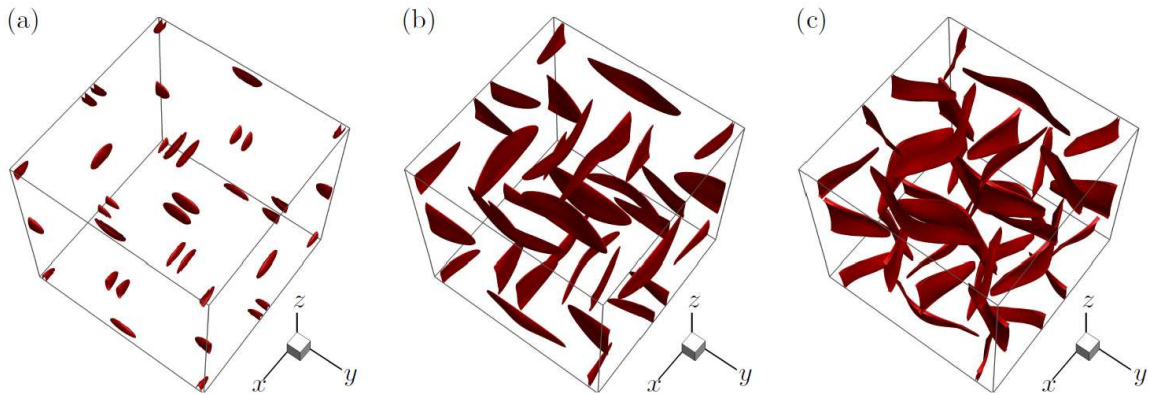}
\caption{Temporal evolution of shocklets at $Ma=2.0$ in the early stage. The dark-red region depicts shocklets defined by $\theta/\theta'<-3$. (a) $t=1$; (b) $t=2$; (c) $t=3$.}\label{fig:shock}
\end{center}
\end{figure}

\begin{figure}
\begin{center}
\includegraphics[width=6.3in]{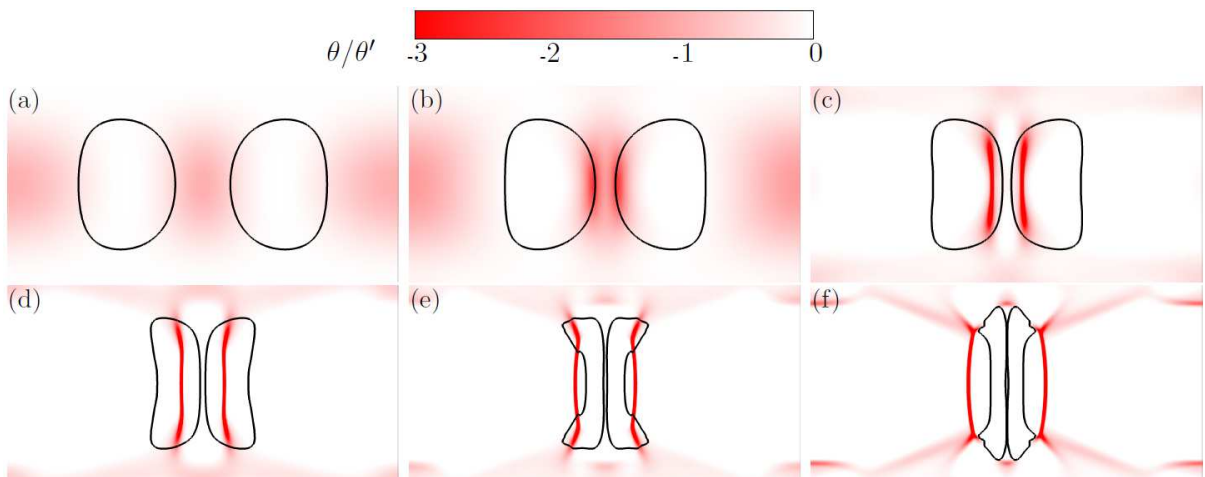}\\[2.0mm]
\caption{Interaction of shocklets and vortex surfaces at $Ma=2.0$ in the early stage on the $x$--$z$ symmetric plane within $0\le x\le 2\pi,~\pi/2\le z\le 3\pi/2$ at $y=\pi$. The contour is color coded by $-3\le\theta/\theta'\le 0$ from red to white. Dark-red regions depict the shocklets with $\theta/\theta'<-3$, and contour lines are the edges of isosurfaces of $\hat{\phi}_v=0.5$. (a) $t=0.5$, (b) $t=1$, (c) $t=1.5$, (d) $t=2$, (e) $t=2.5$, (f) $t=3$.}\label{fig:vas}
\end{center}
\end{figure}

In high-$Ma$ flows, shocklets with finite thickness can be generated and their dynamics are coupled with vortex surfaces, where the shocklets can be defined as the domain with $\theta/\theta'<-3$ \cite{Virk1995,Samtaney2001}.
The evolution of shocklets during~$0\le t\le 3$ at $Ma=2.0$ is shown in Fig.~\ref{fig:shock}. We find that the shocklets gradually expand and become curved.
The interaction of shocklets and vortex surfaces is shown on the symmetric plane within $0\le x\le 2\pi,~\pi/2\le z\le 3\pi/2$ at $y=\pi$ during $0\le t\le 3$ for $Ma=2.0$ in Fig.~\ref{fig:vas}. The dark red region of $\theta/\theta'<-3$ depicts shocklets, and the solid lines represent the boundary of vortex surfaces of $\hat{\phi}_v=0.5$.

As shown in Fig.~\ref{fig:vas}, a pair of shocklets are generated near the symmetric plane at $x=\pi$, and they move in opposite directions during $0\le t\le 3$. Their major dynamics can be explained by a simplified one-dimensional model, which is discussed in detail in Appendix \ref{sec:shocklet_model}.
At the mean time, a pair of vortex surfaces approach to each other, which is qualitatively similar to the motion in incompressible TG flows.
The interaction of shocklets and vortex surfaces becomes notable when the shocklets penetrate into the vortex surfaces in Figs.~\ref{fig:vas}(c) and (d). Subsequently the shocklets move out of the vortex surfaces and further distort the outer vortex edge in Figs.~\ref{fig:vas}(e) and (f). On the other hand, the shocklets are slightly bent by the vortex-induced velocity.

\begin{figure}
\begin{center}
\includegraphics[width=6.2in]{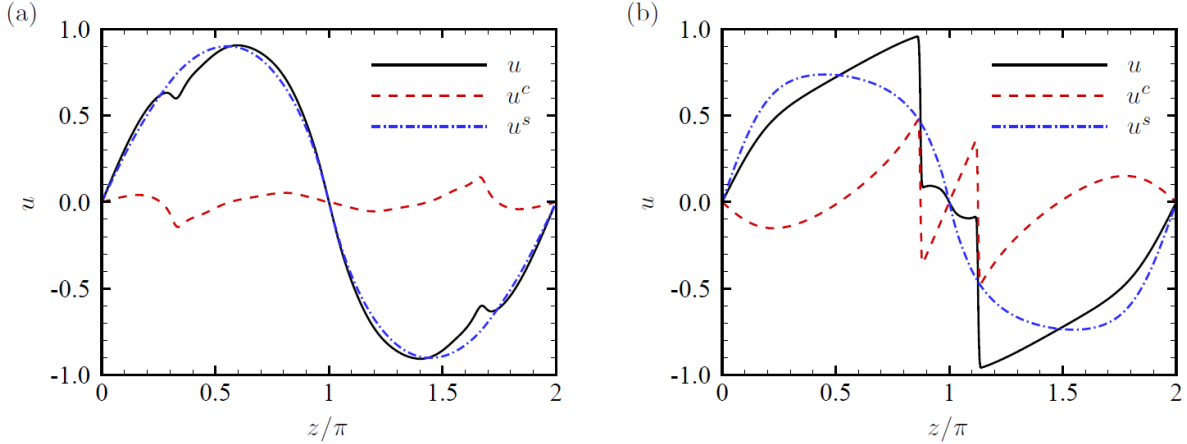}
\caption{Comparison of the velocity component $u$, compressive component $u^c$, and solenoidal component $u^s$ at two Mach numbers on the symmetric line intersected by $x=\pi$ and $y=\pi$ planes at $t=2.2$. (a) $Ma=0.5$, (b) $Ma=2.0$.}\label{fig:ut2.5}
\end{center}
\end{figure}

In order to elucidate the deformation of vortex surfaces, we apply the Helmholtz decomposition \cite{Batchelor1967} to the velocity field as
\EQ
\bm{u}(\bm{x},t)=\bm{u}^c(\bm{x},t)+\bm{u}^s(\bm{x},t),
\EN
where the compressive component $\bm{u}^c$ is irrotational satisfying $\nabla\times\bm{u}^c = \bm{0}$, and the solenoidal component $\bm{u}^s$ satisfies $\nabla\cdot\bm{u}^s = 0$. Fig.~\ref{fig:ut2.5} shows the velocity component $u$ with its compressive and solenoidal components $u^c$ and $u^s$ on the symmetric line intersected by the planes $x=\pi$ and $y=\pi$ at $t=2.5$ for $Ma=0.5$ and 2.0. In Fig.~\ref{fig:ut2.5}(a) for $Ma=0.5$, $u$ is dominated by $u^s$, which is similar to that in incompressible TG flows. In Fig.~\ref{fig:ut2.5}(b) for $Ma=2.0$, the magnitudes of $u^c$ and $u^s$ are comparable near the shocklet represented by the jump of $u^c$ or $u$. The opposite signs of $u^c$ near the jump imply that the shocklet appears to be a velocity sink to contract surrounding vortex surfaces during $1.5\le t\le 2.5$.

\subsection{Vortex reconnection during transition}
\label{sec:results_rec}

\begin{figure}
\begin{center}
\includegraphics[width=6.2in]{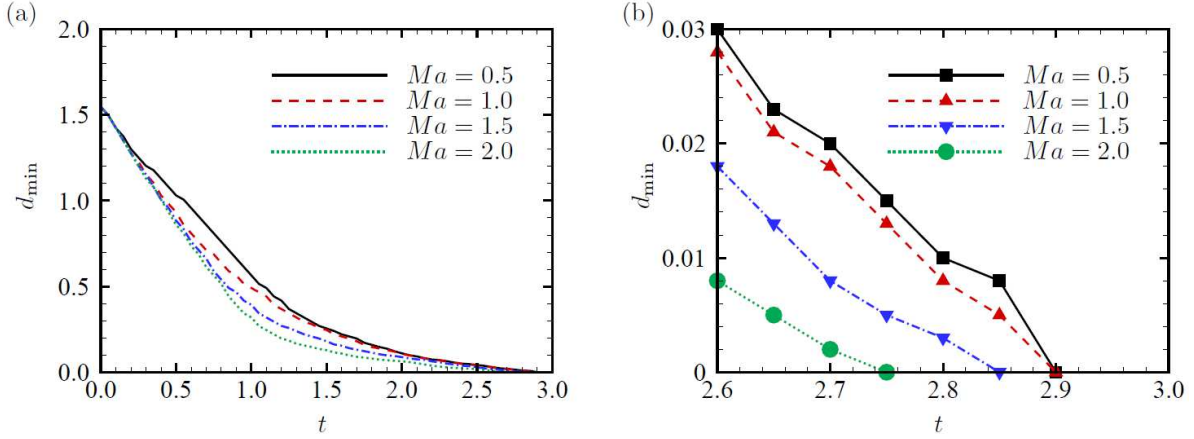}
\caption{Characterization of incipient vortex reconnection at different $Ma$. (a) Temporal evolution of the minimal distance $d_{\min}$ between the pair of VSF isosurfaces of $\hat{\phi}_v=0.5$ at different $Ma$; (b) Zoom-in plot of $d_{\min}$ around the reconnection time.}\label{fig:dmin}
\end{center}
\end{figure}

Vortex reconnection is the topological change of vortex lines or surfaces by cutting and connecting in the vortex evolution \cite{Kida1994}, and it is a critical process in transitional flows \cite{Yang2011,Zhao2016b}. In the VSF evolution, the end of the approaching vortex pair in Fig.~\ref{fig:vas} signals the incipient vortex reconnection. To quantify the reconnection, we plot the temporal evolution of the minimal distance $d_{\min}$ \cite{Zhao2016b} between the approaching vortex surfaces of $\hat{\phi}_v=0.5$ for different $Ma$ in Fig.~\ref{fig:dmin}. We find that $d_{\min}$ decreases as the approaching vortex motion shown in Fig.~\ref{fig:vas}, and $d_{\min}=0$ occurs at the reconnection time $t^*=2.75\sim2.9$ for different $Ma$ when the pair of vortex surfaces merges. Moreover, we find that the reconnection occurs earlier with smaller $t^*$ for larger $Ma$ in Fig.~\ref{fig:dmin}(b), which appears to be contributed to the additional deformation of vortex surfaces induced by shocklets.

\begin{figure}
\begin{center}
\includegraphics[width=6.2in]{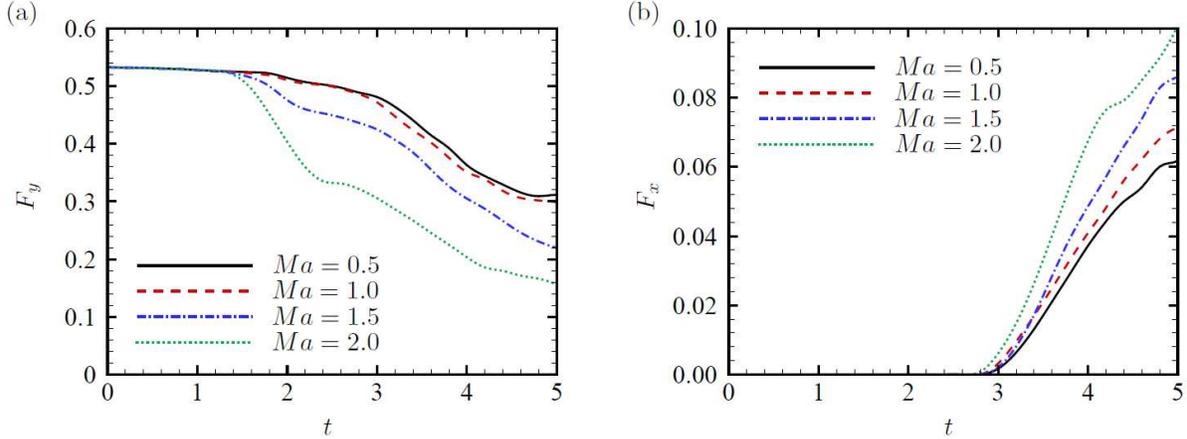}
\caption{Temporal evolution of vorticity fluxes through symmetric planes. (a) $F_y$; (b) $F_x$.}\label{fig:flux}
\end{center}
\end{figure}

Another metric of vortex reconnection is the vorticity flux through the cross section of a closed vortex surface \cite{Zhao2016b}. Without loss of generality, we consider the vorticity flux across the $x$--$z$ symmetric plane at $y=\pi$ between a pair of merging VSF isosurfaces of $\hat{\phi}_v=0.5$. During the reconnection of vortex surfaces, vortex lines are cut in the $y$-direction and reconnected in the $x$-direction (also see Figs.~\ref{fig:2.0phi}(c) and (d)), so the exchange of the vorticity fluxes $F_y\equiv\int_{D_y}\omega_y\rd S$ through the $x$--$z$ plane at $y=\pi$ and $F_x\equiv\int_{D_x}\omega_x\rd S$ through the $y$--$z$ plane at $x=\pi$ quantifies the degree of reconnection, where $D_y$ and $D_x$ are the regions enclosed by VSF isosurfaces of $\hat{\phi}_v=0.5$ on their corresponding symmetric planes. The temporal evolution of $F_y$ and $F_x$ in Fig.~\ref{fig:flux} shows that $F_y$ decreases and $F_x$ increases more rapidly, which implies the higher degree of reconnection, for larger Mach numbers.
The starting times of the significant decrease of $F_y$ and increase of $F_x$ agree with the reconnection time determined by $d_{\min}$ for low $Ma=0.5$ and 1.0.
On the other hand, $F_y$ drops sharply at earlier times than $t^* = 2.75\sim2.85$ for high $Ma=1.5$ and 2.0 owing to the dilatational dissipation and the additional reconnection of vortex lines on the vortex surface in strongly compressible flows.

\begin{figure}
\begin{center}
\includegraphics[width=6.2in]{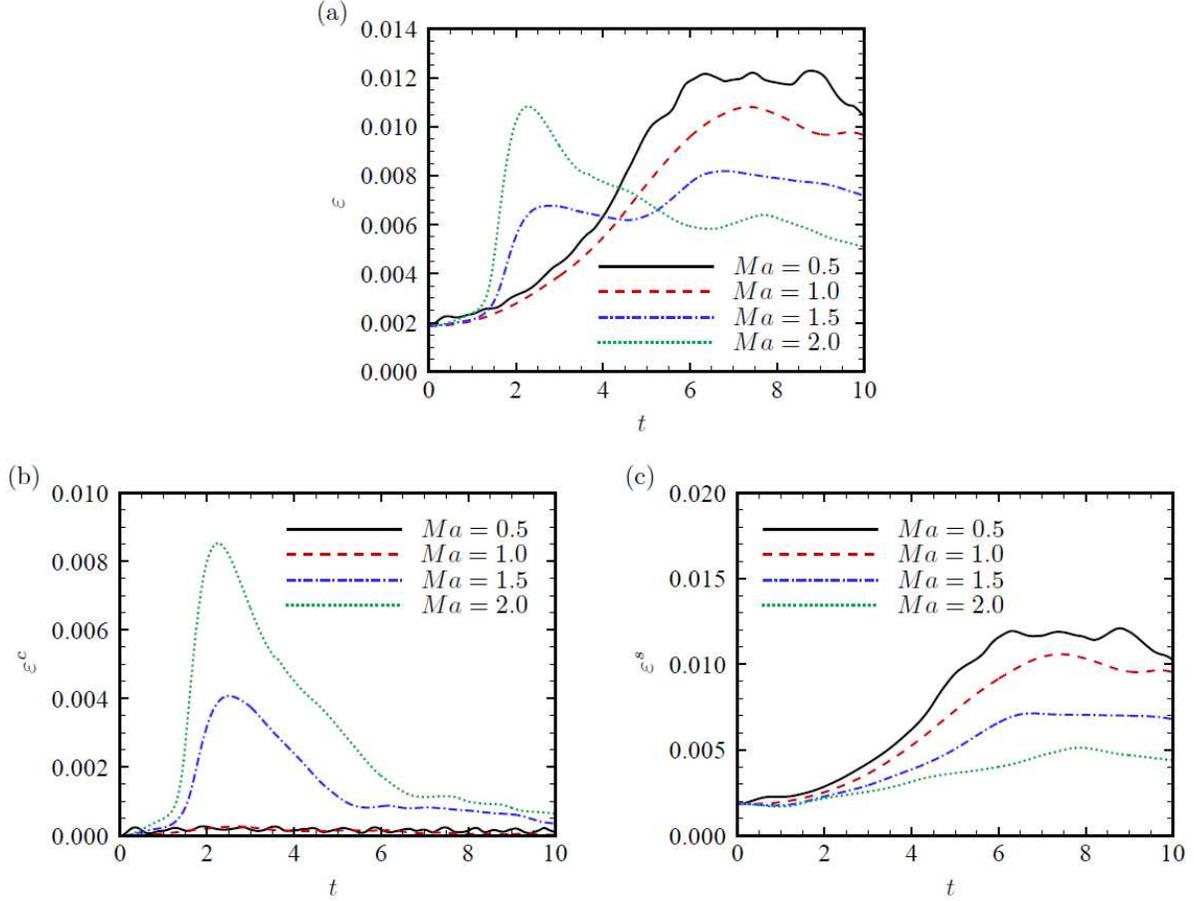}
\caption{Temporal evolution of the dissipation rate and its two components at different $Ma$. (a) Dissipation rate $\varepsilon$; (b) dilatational dissipation $\varepsilon^c$; (c) solenoidal dissipation $\varepsilon^s$.}\label{fig:epo}
\end{center}
\end{figure}

The temporal evolution of the dissipation rate $\varepsilon$ defined by Eq.~\eqref{eq:epsilon} at different $Ma$ is plotted in Fig.~\ref{fig:epo}, as well as its two components, namely the dilatational dissipation $\varepsilon^c\equiv\frac{4}{3}\<\mu\theta^2\>/Re$ and solenoidal dissipation $\varepsilon^s\equiv\<\mu|\bomg|^2\>/Re$.
The dissipation rate reaches a plateau around $t=6$ for $Ma\le 1$ in Fig.~\ref{fig:epo}(a), which is consistent with that in the incompressible TG flow at the same $Re$ \cite{Yang2011}, and the peak value of $\varepsilon$ decreases with increasing $Ma$. A secondary peak of $\varepsilon$ occurs at $t\approx 2.2$ for $Ma>1$, and the peak value increases with increasing $Ma$. The major contribution to the secondary peak for $Ma>1$ is from $\varepsilon^c$ suggested in Figs.~\ref{fig:epo}(b) and (c). The occurrence of the maximum dilatational dissipation indicates the strongest compressible effects with shocklets as discussed in Sec.~\ref{sec:results_early}.
Moreover, $\varepsilon^s$ with a single peak after $t=6$ decreases overall with increasing $Ma$. Thus the strong dilatational dissipation generated by shocklets in the early stage results in the sharp decrease of the magnitude of $F_y$ at $Ma=1.5$ and 2.0 before the reconnection time $t^*$ in Fig.~\ref{fig:flux}(a).

\begin{figure}
\begin{center}
\includegraphics[width=5.0in]{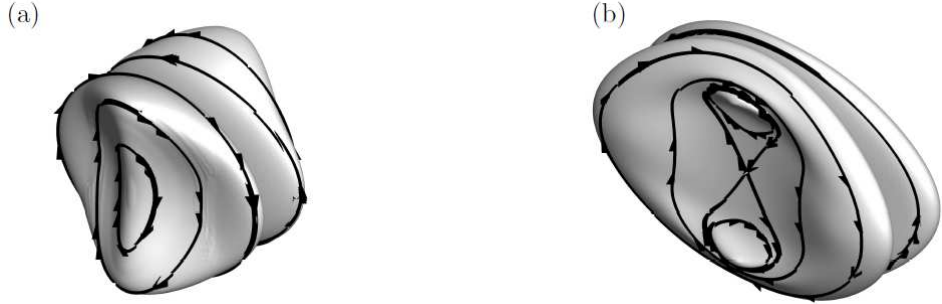}
\caption{Reconnection of vortex lines on the vortex surface at $Ma=2.0$. (a) Before reconnection at $t=2.0$ ; (b) During reconnection at $t=3.0$.}\label{fig:rec2}
\end{center}
\end{figure}

Besides the reconnection of approaching vortex surfaces in Fig.~\ref{fig:rec2}(a), we find the reconnection of vortex lines occurring on the vortex surface at $Ma=2.0$ in Fig.~\ref{fig:rec2}(b). The slight double arch-shaped discontinuity in the middle of the vortex surface is generated by the shocklets in Fig.~\ref{fig:def}(a). As the shocklets move out of the vortex surface as shown in Fig.~\ref{fig:vas}(e), the discontinuity shrinks to two circles in Fig.~\ref{fig:rec2}(b), and a critical point of vortex lines forms at the center of the vortex surface.

\begin{figure}
\begin{center}
\includegraphics[width=6.2in]{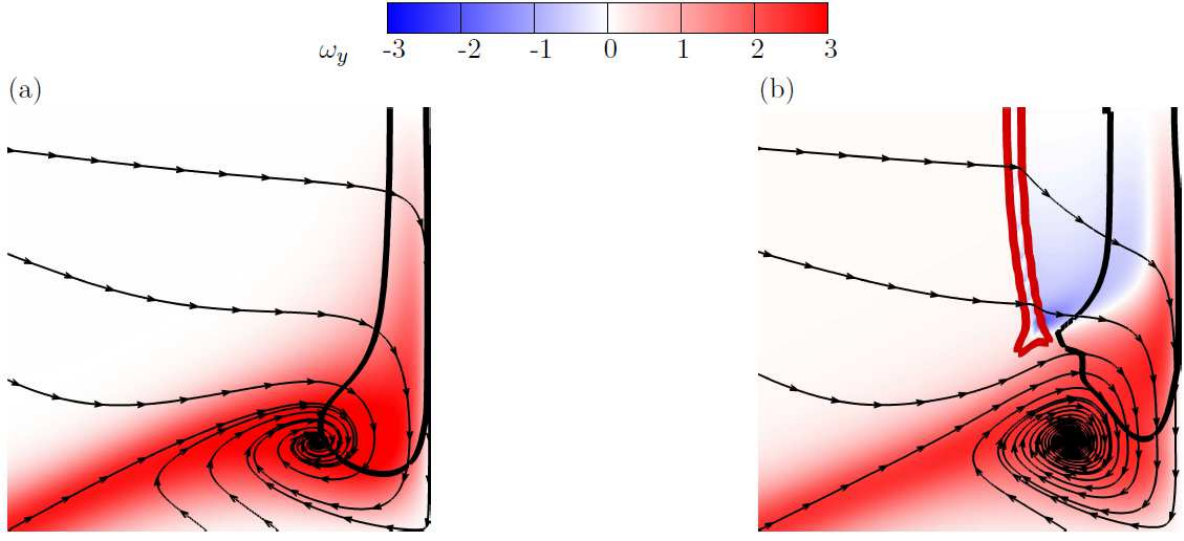}
\caption{Vortex surfaces, shocklets, vorticity component $\omega_y$, and streamlines in the subdomain $\pi/2\le x,y\le\pi$ on the symmetric plane at $y=\pi$ and at $t=3$. (a) $Ma=0.5$; (b) $Ma=2.0$. The contour of $\omega_y$ is color coded by $-3\le\omega_y\le 3$ from blue through white to red, black closed lines are the isocontour lines of $\hat{\phi}_v=0.5$, and the red line in (b) is the isocontour line of $\theta/\theta'=-3$.}\label{fig:recxz}
\end{center}
\end{figure}

We elucidate the mechanism of the reconnection of vortex lines in Fig.~\ref{fig:recxz}, where the black contour lines for the boundaries of the VSF isosurface of $\hat{\phi}_v=0.5$, red contour line for the shocklet boundary with $\theta/\theta'=-3$, contours of the vorticity component $\omega_y$, and streamlines are plotted in the subdomain of $\pi/2\le x,y\le\pi$ on the $x$--$z$ symmetric plane at $y=\pi$ for $Ma=0.5$ and 2.0 at $t=3$. The contour of $\omega_y$ is color coded by $-3\le\omega_y\le 3$ from blue through white to red. For $Ma=0.5$ in Fig.~\ref{fig:recxz}(a), the streamlines are smooth and $\omega_y$ is positive everywhere; for $Ma=2.0$ in Fig.~\ref{fig:recxz}(b), the streamlines crossing the shocklet are kinked owing to the discontinuity of the velocity normal to the shocklet.
This kink changes the concavity or convexity of the streamlines at the edge of vortex surfaces at $Ma=2.0$, which creates negative $\omega_y$ shown by the blue contour in Fig.~\ref{fig:recxz}(b) and leads to the reconnection of vortex lines in Fig.~\ref{fig:rec2}(b). Therefore, the change of signs of the vorticity normal to the $x$--$z$ symmetric plane also diminishes the vorticity flux $F_y$ before $t=t^*$ in Fig.~\ref{fig:flux}(a), though this reconnection of vortex lines does not change the topology of vortex surfaces.


\subsection{Vortex twisting in the late stage}
\label{sec:results_twist}

\begin{figure}
\begin{center}
\includegraphics[width=6.2in]{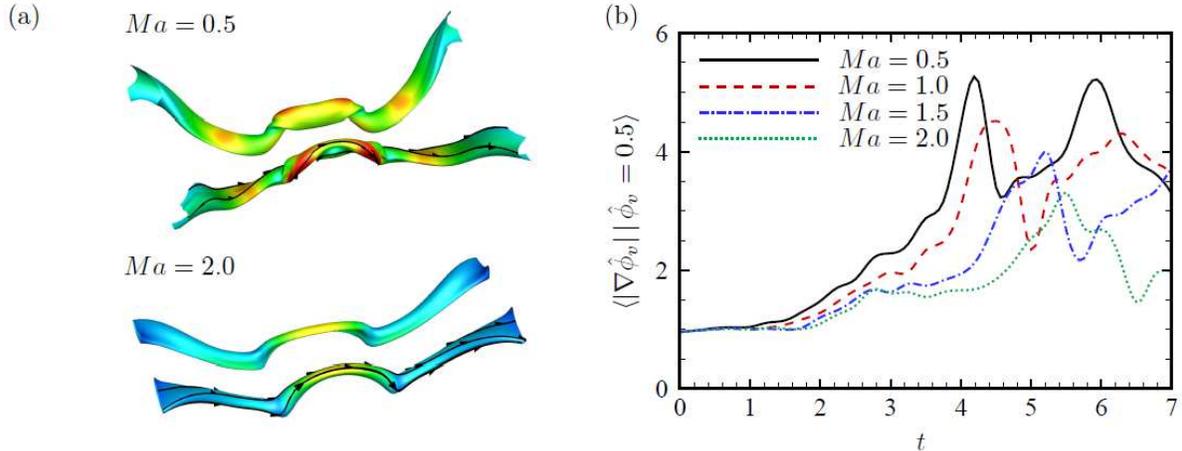}
\caption{Twisting of vortex tubes at different $Ma$. (a) Isosurfaces of $\hat{\phi}_v=0.5$ with $|\bomg|\ge 1$ at $t=7$ for $Ma=0.5$ and 2.0. The surfaces are color coded by $0\le|\bomg|\le 9$ from blue to red; (b) The temporal evolution of  the averaged $|\nabla\hat{\phi}_v|$ conditioned on the isosurface of $\hat{\phi}_v=0.5$ at different $Ma$.}\label{fig:tube}
\end{center}
\end{figure}

After the vortex reconnection, the edge of the merged vortex surface is rolled up into vortex tubes around $t=5$ for moderate and high $Re$, and then the tubes are persistently stretched and twisted in the late stage.
The typical vortex tubes of $\hat{\phi}_v=0.5$ at $Ma=0.5$ and 2.0 at $t=7$ are visualized in Fig.~\ref{fig:tube}(a), where the surfaces with small vorticity $|\bomg|<1$, which have minor influence on flow dynamics, are cut off for clarity.
We find that the geometry of the vortex tubes of $Ma=0.5$ are almost identical to those in the incompressible TG flow \cite{Yang2011}, while the vortex tubes of $Ma=2.0$ are less twisted and have smaller $|\bomg|$ than those in lower-$Ma$ flows.
The averaged magnitude of the scalar gradient $|\nabla\hat{\phi}_v|$ conditioned on the isosurface of $\hat{\phi}_v=0.5$ is used to quantify the deformation of vortex surfaces.
In Fig.~\ref{fig:tube}(b), the temporal evolution of $\<|\nabla\hat{\phi}_v|\,|\,\hat{\phi}_v=0.5\>$ shows that the deformation of vortex surfaces is larger in the early stage at $t<4$ for higher $Ma$ owing to the additional deformation induced by shocklets, while the deformation is generally smaller in the late stage at $t>6$ for higher $Ma$.

\begin{figure}
\begin{center}
\includegraphics[width=6.2in]{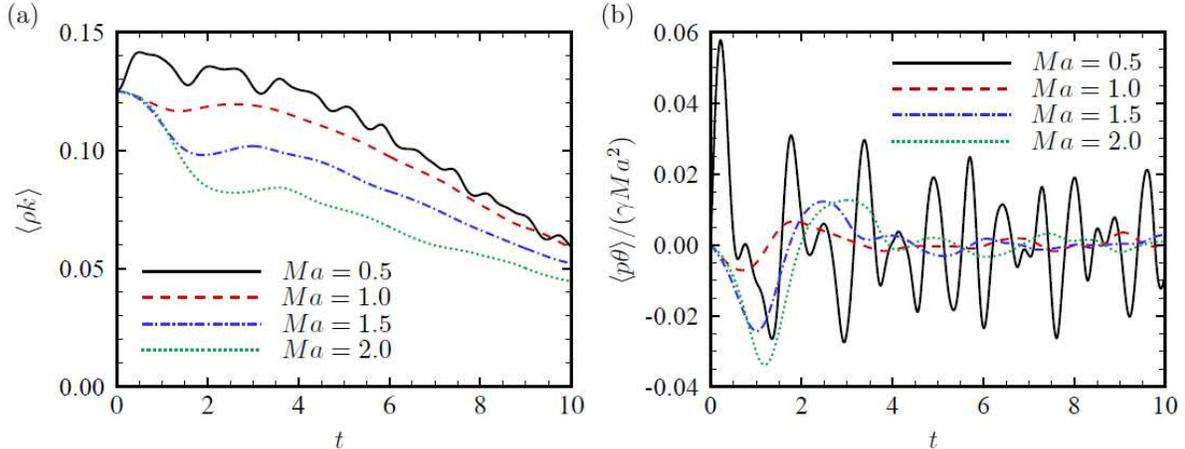}
\caption{The temporal evolution of (a) the average kinetic energy $\<\rho k\>$; (b) the average pressure work $\<p\theta\>/(\gamma Ma^2)$ at different $Ma$.}\label{fig:kpg}
\end{center}
\end{figure}

The mitigation of vortex twisting is consistent with the increasing energy conversion from the kinetic energy to intrinsic energy in high-$Ma$ flows.
Taking the volume average of the energy equation in Eq.~\eqref{eq:nsd} yields the evolution equations of the conversion between the volume averaged kinetic energy $\<\rho k\>$ and intrinsic energy $\<\rho e\>$ as
\EQ
\label{eq:ke}
\CS
\M{}{t}\<\rho k\>=\frac{1}{\gamma Ma^2}\<p\theta\>-\varepsilon,\\
\M{}{t}\<\rho e\>=-\frac{1}{\gamma Ma^2}\<p\theta\>+\varepsilon,
\CN
\EN
where $\<p\theta\>/(\gamma Ma^2)$ is the average pressure work.
The temporal evolution of $\<\rho k\>$ and $\<p\theta\>/(\gamma Ma^2)$ is plotted in Fig.~\ref{fig:kpg}.
Fig.~\ref{fig:kpg}(a) shows that the averaged kinetic energy decreases more rapidly for larger $Ma$, and the fluctuating $\<\rho k\>$ at $Ma=0.5$ is caused by the fluctuating $\<p\theta\>$ in Fig.~\ref{fig:kpg}(b), because the compressibility in weakly compressible flows behaves as a noise source \cite{Moyal1952}.
For $Ma\ge 1.0$ in Fig.~\ref{fig:kpg}(b), $\<p\theta\>/(\gamma Ma^2)$ is negative during the generation of shocklets with $0<t\le 2$, and the negative peak value is comparable with $\varepsilon$ shown in Fig.~\ref{fig:epo}(a).
Then $\<p\theta\>/(\gamma Ma^2)$ converges to zero with small fluctuations around $t>4$ after a slight positive overshoot during $2<t<4$.
Therefore, the positive averaged dissipation rate and the overall negative averaged pressure work accelerate the conversion from kinetic energy to intrinsic energy in Eq.~\eqref{eq:ke} with increasing $Ma$.
The loss of kinetic energy in high-$Ma$ flows cannot sustain the energy cascade to small scales or the dynamical process of twisting vortex tubes.

\section{Conclusions}
\label{sec:conclusion}

We investigate the evolution of VSFs in compressible TG flows at different Mach numbers ranging from $Ma=0.5$ to $2.0$ and at $Re=400$ using DNS.
The evolution equation for VSFs in viscous compressible flows shows that both viscous and baroclinic terms can drive the Lagrangian scalar evolution to deviate from the VSF evolution as the violation of the Helmholtz vorticity theorem,
but the two-time method can effectively reduce the deviation to achieve very small VSF errors $\<|\lambda_{\omega}|\>< 3\%$ in numerical VSF solutions even in high-$Ma$ TG flows.
Furthermore, we develop the renormalization method of VSFs based on the same fluid mass enclosed in VSF isosurfaces to visualize and analyze the continuous evolution of a particular vortex surface.

The effects of the Mach number on the evolution of VSFs are studied in three stages divided by topological and geometrical features of vortex surfaces.
In the early stage, a pair of blob-like vortex surfaces are flattened and a pair of shocklets are generated between the approaching vortex surfaces.
We apply the Helmholtz decomposition to the velocity field and find that the jump of the compressive component of velocity near the shocklet generates a velocity sink to contract surrounding vortex surfaces.
Thus the shocklets moving into the approaching vortex surfaces shrink the vortex volume, and subsequently the shocklets moving out of vortex surfaces distort the edges of vortex surfaces.

Then vortex reconnection occurs when the vortex pair is merged. We use the minimal distance between the approaching vortex surfaces to quantify the reconnection time $t^*$ and find that the reconnection occurs earlier with smaller $t^*$ for larger $Ma$. Another metric of the reconnection time and the reconnection degree is the exchange of vorticity fluxes through two orthogonal cross sections of the approaching closed vortex surfaces. We find that the flux exchange occurs earlier and its magnitude becomes larger with increasing $Ma$, because the shocklets generate the dilatational dissipation and additional reconnection of vortex lines on the vortex surface.

After the vortex reconnection, the edge of vortex surfaces is rolled up into vortex tubes, and the tubes are stretched and twisted in the late stage. We demonstrate that the combination of the positive dissipation rate and overall negative pressure work accelerates the conversion from the kinetic energy to intrinsic energy for large $Ma$, and the loss of kinetic energy in high-$Ma$ flows mitigates the twisting of vortex tubes to suppress the scale cascade in the evolution of vortex surfaces.

The present study demonstrates that the VSF method is applicable to compressible flows. In future work, this method can be applied to investigate the continuous evolution of vortical structures in hydrodynamic instabilities in compressible flows, such as the Richtmyer--Meshkov instability.

\begin{acknowledgments}

Numerical simulations were carried out on the TH-2A supercomputer in Guangzhou, China. This work has been supported in part by the National Natural Science Foundation of China (grant Nos.~11522215 and 11521091), the Science Challenge Project (No.~TZ2016001), and the Thousand Young Talents Program of China.

\end{acknowledgments}



\appendix

\section{Generation and evolution of shocklets in the early stage}\label{sec:shocklet_model}

We develop a one-dimensional model to elucidate the generation and evolution of shocklets in the early stage of compressible TG flows.
Without loss of generality, Eq.~\eqref{eq:ns} on the central symmetric line intersected by $x=\pi$ and $y=\pi$ planes, where $v(x,t)=w(x,t)=0$ from the symmetries in TG flows \cite{Brachet1983}, is simplified as
\EQ
\label{eq:1d}
\CS
\bm{U}_t+\bm{F}_x=\bm{f},\\
\bm{U}(x,t=0)=\bm{U}_0(x)
\CN
\EN
by ignoring the viscous term, where the vector-valued quantities are
\EQ
\bm{U}=
\lrn{\AR{c}
	\rho\\
	\rho u\\
	E
	\AN},~~
\bm{F}=
\lrn{\AR{c}
	\rho u\\
	\rho u^2+\frac{p}{\gamma Ma^2}\\
	\lrn{\rho u^2+\frac{p}{\gamma Ma^2}}u
	\AN},~~
\bm{f}=
\lrn{\AR{c}
	-\rho\theta_{yz}\\
	0\\
	0
	\AN}
\EN
with
\EQ
\theta_{yz}(x,t)=\D{v}{y}(x,t)+\D{w}{z}(x,t).
\EN
Here the subscripts `$t$' and `$x$' denote partial derivatives of $t$ and $x$, respectively. Compared with the one-dimensional Euler equations, the nonhomogeneous term $\bm{f}$ in Eq.~\eqref{eq:1d} represents the three-dimensional effects in the compressible TG flow on the one-dimensional model.

We divide the time period $[0,t]$ into $N_t$ uniform intervals $\delta t=t/N_t=t_{j+1}-t_j$, $j=0,1,2,\cdots,N_t$.
Letting $\bm{U}(x,t_j)$ be the solution of Eq.~\eqref{eq:1d} at $t=t_j$,
we use the operator splitting method \cite{Peaceman1955} to solve Eq.~\eqref{eq:1d} at $t>t_j$.
First ignoring the nonhomogeneous term $\bm{f}$ yields the one-dimensional Euler equations as
\EQ
\label{eq:1Dconserv}
\CS
\bm{U}_t+\bm{F}_x=\bm{0},\\
\bm{U}(x,t=t_j)=\bm{U}(x,t_j).
\CN
\EN
Its solution
\EQ
\label{eq:utilde}
\widetilde{\bm{U}}(x,t_{j+1})=
\lrn{\AR{c}
	\widetilde{\rho}(x,t_{j+1})\\
    \widetilde{\rho}(x,t_{j+1})\widetilde{u}(x,t_{j+1})\\
    \widetilde{E}(x,t_{j+1})
	\AN}
\EN
at $t=t_{j+1}$ can be obtained by numerical calculation.

Next, we add the nonhomogeneous term $\bm{f}$ into Eq.~\eqref{eq:1Dconserv} and update the initial condition as
\EQ
\label{eq:qch}
\CS
\bm{U}_t+\bm{F}_x=\bm{f},\\
\bm{U}(x,t=t_j)=\widetilde{\bm{U}}(x,t_{j+1}).
\CN
\EN
Assuming $\theta_{yz}(x,t)$ is a known source term, which can be extracted from the DNS result,
the solution of Eq.~\eqref{eq:qch} is
\EQ\label{eq:qch_sol}
\bm{U}(x,t_{j+1})=\widetilde{\bm{U}}(x,t_{j+1})\re^{-\theta_{yz}(x,t_j)\delta t}.
\EN
Finally, we obtain the solution of Eq.~\eqref{eq:1d} as
\EQ
\label{eq:1ds}
\bm{U}(x,t)=\lim\limits_{N_t\to\infty}\sum\limits_{j=0}^{N_t}\widetilde{\bm{U}}(x,t_{j+1})\re^{-\theta_{yz}(x,t_j)\delta t}
\EN
by the superposition principle \cite{Evans1998}.
In Eq.~\eqref{eq:1ds}, $\bm{U}(x,t)$ is contributed from the product of $\widetilde{\bm{U}}(x,t)$ and a decaying function $\re^{-\theta_{yz}(x,t)\delta t}$. Compared with the one-dimensional Euler equations, the temporal evolution of Eq.~\eqref{eq:1d} is inhibited in the domain with positive $\theta_{yz}(x,t)$, or enhanced with negative $\theta_{yz}(x,t)$.

The initial condition of Eq.~\eqref{eq:1d} for the compressible TG flow is
\EQ
\label{eq:init}
\CS
u_0(x)=\sin(x),\\
\rho_0(x)=p_0(x)=1+\frac{3}{16}\lrl{1+\cos(2x)}.
\CN
\EN
In the domain of $\pi/2\le x\le 3\pi/2$, $\partial u_0(x)/\partial x<0$ implies that the fluid tends to move towards the center at $x=\pi$ in Eq.~\eqref{eq:1Dconserv}.
This converging motion generates a shock wave at the center $x=\pi$, and then the accumulated pressure and density at the center drive the central shock wave to split into two moving in opposite directions.

We compare the temporal evolutions of the velocity component $u$ calculated from DNS and modeling at $Ma=2.0$ on the symmetric line intersected by $x=\pi$ and $y=\pi$ planes.
Fig.~\ref{eq:1ds}(a) shows the one-dimensional result extracted from DNS;
Fig.~\ref{eq:1ds}(b) shows the solution Eq.~\eqref{eq:1ds} of the one-dimensional model Eq.~\eqref{eq:1d} with the initial condition Eq.~\eqref{eq:init} and the source term $\theta_{yz}$ extracted from DNS;
Fig.~\ref{eq:1ds}(c) shows the numerical solution of the one-dimensional Euler Eq.~\eqref{eq:1Dconserv} with the initial condition Eq.~\eqref{eq:init}.

\begin{figure}
\begin{center}
\includegraphics[width=6.2in]{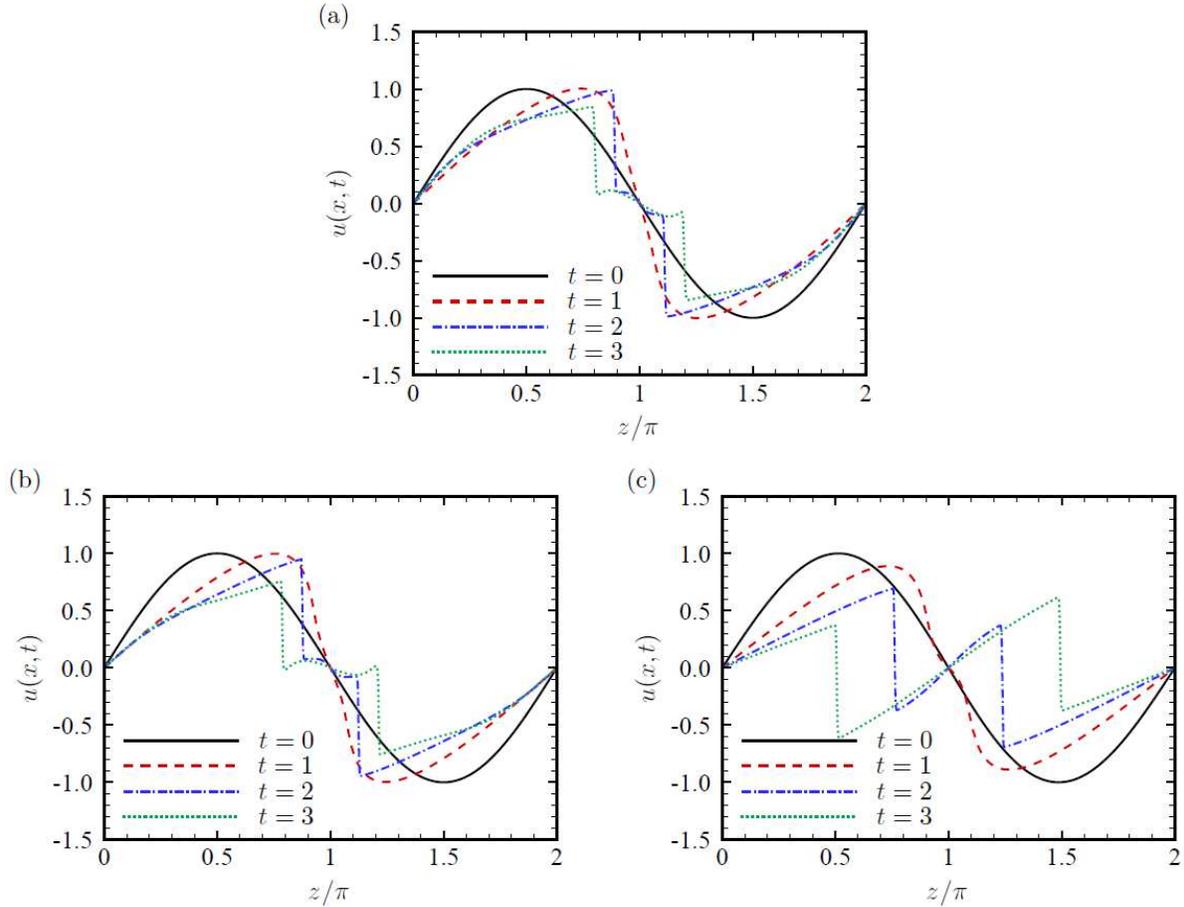}
\caption{Temporal evolutions of the velocity component $u$ calculated from DNS and modeling on the symmetric line intersected by $x=\pi$ and $y=\pi$ planes at $Ma=2.0$. (a) The one-dimensional result extracted from DNS; (b) the solution Eq.~\eqref{eq:1ds} of the one-dimensional model Eq.~\eqref{eq:1d} with the initial condition Eq.~\eqref{eq:init} and the source term $\theta_{yz}$ extracted from DNS; (c) the solution of the one-dimensional Euler Eq.~\eqref{eq:1Dconserv} with the initial condition Eq.~\eqref{eq:init}.}
\label{fig:u}
\end{center}
\end{figure}

We find that the solution Eq.~\eqref{eq:1ds} is a very good approximation to the DNS result from the good agreement of the $u$ profiles in Figs.~\ref{fig:u}(a) and (b), so the model Eq.~\eqref{eq:1d} captures the major one-dimensional dynamics in the generation of shocklets.
Furthermore, the generation and moving direction of shocklets are similar to those of the splitted shock waves calculated from the one-dimensional Euler Eq.~\eqref{eq:1Dconserv} in Fig.~\ref{fig:u}(c).
On the other hand, the fluid near the shocklets moving along the $x$-direction tends to expand in the $y$--$z$ planes in the compressible TG flows.
Compared with the one-dimensional shock waves in Fig.~\ref{fig:u}(c), this three-dimensional process generates positive $\theta_{yz}(x,t)$ to smooth the shocklets and inhibit their evolution as shown in Figs.~\ref{fig:u}(a) and (b) and implied in Eq.~\eqref{eq:qch_sol}.

\bibliographystyle{elsarticle-num}
\bibliography{VSFCTG}

\end{document}